# AS QUESTÕES DE FÍSICA E O DESEMPENHO DOS ESTUDANTES NO ENEM
# PHYSICS ITEMS AND STUDENT'S PERFORMANCE AT ENEM

**Wanderley P. Gonçalves Jr[1], Marta F. Barroso[2]**
[1]Universidade Federal do Rio de Janeiro / Instituto de Física e Colégio de Aplicação, wpgjunior@if.ufrj.br
[2]Universidade Federal do Rio de Janeiro / Instituto de Física, marta@if.ufrj.br

**Resumo**
O Exame Nacional do Ensino Médio (ENEM) passou por uma reformulação no ano de 2009, modificando seus objetivos: de uma auto-avaliação de competências ao final do ensino médio para uma avaliação que possibilita o acesso às universidades e ao financiamento estudantil. Nesta mudança, de uma única prova passa-se a quatro provas por grandes áreas, constrói-se uma Matriz de Referência com domínios cognitivos, competências, habilidades e objetos de conhecimento, e muda-se também a metodologia de análise, com a utilização da Teoria da Resposta ao Item, que possibilita a comparação longitudinal dos resultados de anos diferentes, com a intenção de monitorar o ensino médio no país. Neste trabalho, apresentamos um estudo sobre as questões de Física da prova de Ciências da Natureza do ENEM 2009, 2010 e 2011, posteriores à mudança. A partir de categorias definidas, são propostas variáveis qualitativas para caracterizar o tipo de prova. Também são analisadas as respostas dos estudantes participantes da prova de Ciências da Natureza em 2009, usando dados disponibilizados pelo INEP. A análise qualitativa revela as características da prova nestes anos: questões longas, com pouca exigência de raciocínios mais complexos característicos da resolução de problemas, e uma tendência de distribuição de questões por objetos de conhecimento diferente do tradicional no ensino médio. A análise do desempenho dos alunos também é reveladora dos processos efetivos de aprendizagem da disciplina, indicando que o percentual de acertos nos itens quase sempre é baixo, e que questões que exigem algum tipo de conhecimento disciplinar ou que exigem utilização de raciocínios matemáticos apresentam um desempenho sensivelmente mais fraco.

**Abstract**
The Brazilian National Assessment of Secondary Education (ENEM, Exame Nacional do Ensino Médio) has changed in 2009: from a self-assessment of competences at the end of high school to an assessment that allows access to college and student financing. From a single general exam, now there are tests in four areas: Mathematics, Language, Natural Sciences and Social Sciences. A new Reference Matrix is build with components as cognitive domains, competencies, skills and knowledge objects; also, the methodological framework has changed, using now Item Response Theory to provide scores and allowing longitudinal comparison of results from different years, providing conditions for monitoring high school quality in Brazil. We present a study on the issues discussed in Natural Science Test of ENEM over the years 2009, 2010 and 2011. Qualitative variables are proposed to characterize the items, and data from students´ responses in Physics items were analysed. The qualitative analysis reveals the characteristics of the exam in these years: long items, only a few of them demanding more complex reasoning, a characteristic of problem solving skills. The analysis of student performance also reveals that learning physics is not attained, with a percentage of correct answers on items that is almost always small, and that items that require some type of disciplinary knowledge or require use of mathematical reasoning presents a performance significantly weaker.

## Introdução

O Exame Nacional do Ensino Médio (ENEM) foi concebido com o objetivo de "*avaliar o indivíduo ao término da escolaridade básica, para aferir o desenvolvimento de competências fundamentais ao exercício pleno da cidadania*", conforme o Documento Básico do ENEM [1] elaborado pelo Instituto Nacional de Estudos e Pesquisas Educacionais Anísio Teixeira (INEP), órgão do Ministério da Educação.

No período de 1998 a 2008, a avaliação foi feita por meio da aplicação de um exame único, com 63 questões objetivas de múltipla escolha e uma redação. O desempenho dos participantes era mensurado pela nota da prova objetiva, cujo valor correspondia ao percentual de questões respondidas corretamente (num total de 0 a 100), e pela nota da redação, também valendo 100 pontos. Em vez de um programa para este exame, foi elaborada uma Matriz de Referência, termo que se refere às múltiplas dimensões a serem avaliadas simultaneamente pelas questões, algumas delas envolvendo conceitos abstratos, como exemplo, as competências dos examinandos, as habilidades desenvolvidas por eles e os conteúdos aprendidos. A Matriz de Referência para o ENEM, segundo o Documento Básico [1], pressupunha "*a colaboração, a complementaridade e integração entre os conteúdos das diversas áreas do conhecimento presentes nas propostas curriculares das escolas brasileiras de ensino fundamental e médio*", e considerava "*que conhecer é construir e reconstruir significados continuamente, mediante o estabelecimento de relações de múltipla natureza individuais e sociais*".

Neste contexto, o ENEM fornecia um indicador da qualidade do sistema, mas não uma mensuração precisa desta qualidade e, por apresentar características incomuns para uma avaliação em larga escala [2], como não ser um exame obrigatório e não utilizar mecanismos de amostragem, seus resultados possibilitavam a avaliação do desempenho individual dos candidatos e, eventualmente, forneciam um indicador para as escolas que possuíam mais de 90% de seus concluintes participando do exame.

Aos poucos, os resultados do ENEM também passaram a servir como indicadores para financiamento de cursos superiores (Programa Universidade para Todos – PROUNI, a partir de 2005) e para ingresso em universidades, tornando-se assim um exame cada vez mais importante para os estudantes no país.

No ano de 2009, ocorreu uma reformulação no ENEM. Metodologicamente, sua Matriz de Referência foi substituída e as notas dos estudantes passaram a ser escores calculados usando-se a Teoria da Resposta ao Item. Os objetivos do Novo ENEM modificaram-se: além de seus resultados servirem de referência para a auto-avaliação do candidato, eles também passaram a compor os mecanismos de acesso a programas governamentais e aos cursos profissionalizantes, pós-médios e à Educação Superior. Passaram a fornecer, ainda, a certificação de conclusão do ensino médio para jovens e adultos e a ser utilizados na avaliação do desempenho acadêmico dos ingressantes nas Instituições de Ensino Superior (como componente dos resultados do Exame Nacional de Desempenho de Estudantes, o ENADE, no Sistema Nacional de Avaliação do Ensino Superior, SINAES).

Em relação a seu formato, o exame passou a se constituir de 4 provas com 45 questões de múltipla escolha cada, nas áreas de Ciências Humanas, Ciências da Natureza, Linguagens e Códigos e Matemática e suas Tecnologias além de uma prova de Redação. O candidato dispõe de 4 horas e meia, num primeiro dia, para resolver as 90 questões das provas de Ciências da Natureza e Ciências Humanas, e de 5 horas e meia num segundo dia para as 90 questões da prova de Linguagens e Matemática e para a Prova de Redação.

A Matriz de Referência passou a ser composta por 5 eixos cognitivos, comuns a todas as áreas (dominar linguagens, compreender fenômenos, enfrentar situações-problema, construir argumentação, elaborar propostas), com uma Matriz específica para cada uma das áreas, apresentando nessa uma lista de competências e habilidades.

Essas reformulações do exame fizeram com que o ENEM se tornasse uma avaliação muito procurada pelos estudantes que estão concluindo o ensino médio. Já era um exame bastante procurado desde o início da utilização de seus resultados no programa PROUNI, de concessão de bolsas para o sistema de ensino superior privado; em 2011, quando muitas instituições federais de ensino superior aderiram ao Sistema de Seleção Unificada (SiSU), o número de inscritos superou 5 milhões.

Tudo isso faz com que o ENEM forneça indicadores relevantes sobre a qualidade da educação básica brasileira, mesmo com seus vícios de origem: ele não foi proposto para ser uma avaliação de sistema, e sim para ser uma avaliação individual e, principalmente, o exame não é universal nem amostral para os estudantes do ensino médio.

Neste trabalho, apresenta-se um estudo sobre aspectos qualitativos das questões de Física do ENEM dos anos de 2009, 2010 e 2011 [3]. Com base em um conjunto de variáveis qualitativas construídas a partir das características do ensino de Física no nível médio, todas as questões de Física foram classificadas. Os microdados do exame de 2009 foram disponibilizados pelo INEP, e análise destes dados permite observar as dificuldades dos estudantes que prestaram o exame. Em particular, foi possível relacionar os aspectos qualitativos das questões com o desempenho dos alunos. Observa-se por exemplo que as questões que envolvem qualquer tipo de manipulação matemática apresentam um percentual de acerto mais baixo do que os itens estritamente qualitativos. Com base neste estudo, é possível apresentar algumas conclusões a respeito do que o exame revela em relação à aprendizagem em Física dos estudantes ao final do ensino médio, e sobre as possíveis repercussões no ensino e aprendizagem de física e ciências no ensino médio proporcionado pela utilização maciça deste exame para ingresso ou financiamento nas instituições de ensino superior.

## Avaliações em Larga Escala

A avaliação é uma peça fundamental no ensino e aprendizagem; no entanto, os professores em geral simplesmente reproduzem os processos avaliativos que conheceram em sua formação, refletindo muito pouco sobre estes.

Enquanto as avaliações se restringiam às paredes de sala de aula, cada professor aplicava-as e interpretava seus resultados de forma individual ou no máximo dentro de sua instituição de ensino. A partir da década de 90, com o Sistema de Avaliação da Educação Básica (SAEB), em 1998 com o ENEM, e com o SINAES (ENADE) na década de 2000, o país começou a seguir uma tendência internacional [2] de realização de processos de avaliação externa da aprendizagem dos alunos e da qualidade do ensino. Esta política passou a gerar polêmicas e preocupações, ao possibilitar a criação de políticas de responsabilização [4] e comparação entre escolas. Tal preocupação aumentou no ensino médio após as mudanças ocorridas em 2009 no ENEM.

Dentro deste contexto, torna-se necessária uma reflexão em relação à avaliação, em especial aos métodos em larga escala, para que seus resultados sejam compreendidos e, principalmente, possam ser utilizados como agentes de melhoria da aprendizagem.

As avaliações em larga escala constituem um tipo de avaliação educacional que têm como objetivo a obtenção de dados e a realização de análises que permitam diagnosticar a situação da educação, fornecendo subsídios para a implementação, a manutenção e a reformulação de políticas educacionais. O termo “larga escala” refere-se a um número muito grande de testes aplicados ou à utilização de algum tipo de amostragem estatística. Uma das características desse tipo de avaliação é que elas possibilitam o monitoramento contínuo dos processos educativos, permitindo detectar tanto os benefícios quanto os malefícios decorrentes de políticas educacionais adotadas [5]; e, em geral, essas avaliações não se destinam a analisar e fornecer dados de alunos ou escolas individualmente [2].

A partir da década de 90, o número de avaliações em larga escala cresce substancialmente em todo o mundo [2; 6]. Essas avaliações são focadas, quase sempre, em Linguagens, Matemática e Ciências. Alguns exemplos internacionais são o PISA (Programme for International Student Assessment), o TIMSS (Trends in Mathematics and Science Study), e um exemplo nacional é o SAEB. O ENEM é uma avaliação em larga escala pelo número de estudantes que presta o exame: como a realização do exame é opcional para o estudante, ele não se constitui numa avaliação universal nem em uma avaliação amostral, exigindo portanto um certo cuidado nas extrapolações de seus resultados para a análise dos sistemas de ensino.

A compreensão e entendimento desse tipo de avaliação tornam-se fundamentais, principalmente para os professores, a partir do momento que, ao servir de base para as ações das políticas públicas em educação, passam a determinar, direta e indiretamente, o currículo a ser ensinado nas escolas, as cargas horárias das disciplinas e, finalmente, o perfil dos alunos que ingressam nas universidades.

**O Exame Nacional do Ensino Médio**

O ENEM é um exame com questões, denominadas itens, de múltipla escolha, elaboradas segundo a concepção [7] de que a questão deve conter um texto base, um enunciado e as alternativas. O texto base deve apresentar as informações necessárias para a solução da situação problema proposta; o enunciado oferece uma formulação objetiva da tarefa a ser realizada; e as alternativas são as possibilidades de resposta apresentadas. Este modelo, adotado pelo INEP, é um dos muito possíveis na elaboração de questões de múltipla escolha [8] e a princípio fornece informações relevantes sobre os objetivos do processo avaliativo.

Os itens são elaborados a partir de uma Matriz de Referência [9] que contém quatro dimensões: domínios cognitivos, competências, habilidades e objetos de conhecimento. Os eixos cognitivos são 5, comuns a todas as áreas: dominar linguagens, compreender fenômenos, enfrentar situações-problema, construir argumentação e elaborar propostas. As competências são diferentes em cada área; as da área de Ciências da Natureza são 8, e estão apresentadas no Quadro 1.

**Quadro 1.** As competências da área de Ciências da Natureza da Matriz de Referência do ENEM

| |
|---|
| *Competência 1 – Compreender as ciências naturais e as tecnologias a elas associadas como construções humanas, percebendo seus papéis nos processos de produção e no desenvolvimento econômico e social da humanidade.* |
| *Competência 2 – Identificar a presença e aplicar as tecnologias associadas às ciências naturais em diferentes contextos.* |
| *Competência 3 – Associar intervenções que resultam em degradação ou conservação ambiental a processos produtivos e sociais e a instrumentos ou ações científico-tecnológicos.* |
| *Competência 4 – Compreender interações entre organismos e ambiente, em particular aquelas relacionadas à saúde humana, relacionando conhecimentos científicos, aspectos culturais e características individuais.* |
| *Competência 5 – Entender métodos e procedimentos próprios das ciências naturais e aplicá-los em diferentes contextos.* |
| *Competência 6 – Apropriar-se de conhecimentos da física para, em situações problema, interpretar, avaliar ou planejar intervenções científico-tecnológicas.* |
| *Competência 7 – Apropriar-se de conhecimentos da química para, em situações problema, interpretar, avaliar ou planejar intervenções científico-tecnológicas.* |
| *Competência 8 – Apropriar-se de conhecimentos da biologia para, em situações problema, interpretar, avaliar ou planejar intervenções científico-tecnológicas.* |

A cada uma dessas competências, são associadas habilidades, que totalizam 30 na área de Ciências da Natureza.

Os objetos de conhecimento são divididos disciplinarmente. Na área de Física, são associados em 7 grandes áreas, apresentadas no Quadro 2.

**Quadro 2**. Os objetos de conhecimento em Física

1. *Conhecimentos básicos e fundamentais;*
2. *O movimento, o equilíbrio e a descoberta de leis físicas;*
3. *Energia, trabalho e potência;*
4. *A mecânica e o funcionamento do universo;*
5. *Fenômenos elétricos e magnéticos;*
6. *Oscilações, ondas, óptica e radiação;*
7. *O calor e fenômenos térmicos.*

O desempenho do aluno não é medido por uma nota, como na denominada Teoria Clássica dos Testes, quando a nota é apenas a soma (ou média ponderada) dos acertos nas questões. A metodologia de obtenção dos resultados fornece escores individuais por meio da utilização da Teoria da Resposta ao Item [10], que possibilita a criação de uma escala correspondente à aptidão do estudante que pode ser comparada longitudinalmente (em anos sucessivos). A adoção desta metodologia, mais robusta do ponto de vista de atribuição de uma medida à aprendizagem dos estudantes, exige um número grande de itens em cada uma das provas para que o resultado seja confiável. A construção da prova aplicada aos estudantes deve levar em conta as múltiplas dimensões da Matriz de Referência (habilidades, competências, disciplina e objetos do conhecimento, além dos eixos cognitivos).

**A Metodologia de Análise**

As provas (enunciado dos itens) e os microdados (as informações com relação ao padrão de preenchimento das questões da prova e dos dados sócio-econômicos, sem possibilidade de identificação do respondente) são disponibilizadas na página do INEP. No entanto, apenas recentemente (janeiro de 2012) os microdados do ENEM 2009 foram divulgados.

O processo de análise desses dados é importante para que esses resultados possam ser utilizados pelos professores, para compreensão da aprendizagem dos estudantes ao final do ensino médio. Essa análise pode ser realizada em múltiplas dimensões e perspectivas. Neste trabalho, pretendeu-se fazer uma primeira verificação do que esses resultados revelam a respeito da aprendizagem de física no ensino médio no Brasil.

A análise qualitativa de questões de múltipla escolha permite comparar a relação entre o que é o objetivo explícito do processo de avaliação e o que efetivamente é mensurado. Para esta análise qualitativa, considerou-se que o exame possui uma Matriz de Referência pública [9], contendo domínios cognitivos, competências, habilidades e objetos de conhecimento. Portanto, uma primeira categoria [11] foi a adequação das questões ao objetivo avaliativo proposto.

Foram selecionadas as provas de Ciências da Natureza do Exame Nacional do Ensino Médio a partir do ano de sua reformulação. São cinco as provas: 2009 (a primeira versão, não aplicada devido ao vazamento dentro da gráfica, e a segunda versão, aplicada), a de 2010 (a primeira e segunda aplicações; a segunda aplicação foi feita por defeitos de impressão da prova) e a de 2011.

Fez-se a análise das questões de Física dessas provas [6; 12]. As “questões de física” são as que abordam diretamente o conteúdo específico disciplinar de Física. No caso de questões que apresentam traços inter, multi ou transdisciplinares, a identificação foi feita de acordo com o conhecimento exigido para identificar a alternativa correta do item.

A partir da seleção disciplinar, escolheram-se variáveis qualitativas dentro das categorias propostas para o estudo (adequação aos objetivos propostos nos documentos oficiais, tipos de raciocínio utilizados e adequação ao conteúdo disciplinar de física tradicional no ensino médio): o domínio da habilidade avaliada no item (segundo a classificação proposta pelo INEP); o tamanho dos textos; a existência ou não de objetos visuais (gráficos, tabelas, entre outros); o nível de contextualização; a exigência de conhecimentos específicos disciplinares de Física para resolução; a classificação da necessidade de cálculos quantitativos ou não para a resolução do item.

Os resultados da análise dessas variáveis foram validados qualitativamente. Para isso, a classificação das questões nas variáveis foi feita individualmente por mais de um pesquisador, com a comparação dos resultados. No caso de discrepâncias na classificação, era feita uma discussão coletiva até uma conclusão consensual.

Cada uma das questões foi classificada segundo as competências e habilidades exigidas para sua resolução. Em caso de dúvida entre duas ou mais habilidades para o mesmo item, identificou-se dentre elas qual, na Matriz de Referência, estava associada à competência mais adequada. Se ainda restasse dúvida entre duas habilidades dentro de uma mesma competência, classificava-se o item na habilidade ainda não contemplada.

Em seguida, foi feita a classificação dos conteúdos abordados nos itens utilizando a lista dos objetos de conhecimento associados à Matriz de Referência de Ciências da Natureza.

Outras categorias utilizadas relacionaram-se ao tempo de reflexão na realização do exame, que impacta nos resultados; o tipo de contextualização [13] proposto no item; a utilização de processos cognitivos de níveis mais elevados [14], com a utilização de raciocínios abstratos. Para todas essas categorias, foram buscadas variáveis que permitissem a construção de um indicador operacional do conceito a ser empiricamente observado [15].

Para a variável que mede o tempo de reflexão na resolução da questão, utilizou-se uma medida do tamanho das questões, a partir do número de linhas do texto-base e dos enunciados das questões. Para isso, considerou-se a formatação original da prova, com as páginas do caderno de questões divididas em duas colunas.

O nível de contextualização do item foi classificado segundo Nentwig [13], que pressupõe que um item tem um alto nível de contextualização se no texto estão presentes as informações relevantes para sua solução, isto é, se a extração e processamento da informação contida no texto é necessária para a resolução do item; por outro lado, um baixo nível de contextualização está presente se a informação fornecida no texto não é essencial para responder ao item. Adicionou-se a proposta de classificação como pré-texto [16] para a situação em que o texto é completamente desnecessário para a solução do item.

A necessidade de conhecimento específico de física para a solução do item também foi avaliada: classificou-se como inexistente quando a resposta está contida no texto do item, ou existente quando é necessária alguma informação específica não contida no texto. Um exemplo de item que não exige conhecimento específico de física está indicado no Quadro 3, correspondendo à questão 63 da prova Azul de 2011.

**Quadro 3.** Questão 63, Prova Azul, 2011 – sem exigência de conhecimento específico

**QUESTÃO 63**

Para que uma substância seja colorida ela deve absorver luz na região do visível. Quando uma amostra absorve luz visível, a cor que percebemos é a soma das cores restantes que são refletidas ou transmitidas pelo objeto. A Figura 1 mostra o espectro de absorção para uma substância e é possível observar que há um comprimento de onda em que a intensidade de absorção é máxima. Um observador pode prever a cor dessa substância pelo uso da roda de cores (Figura 2): o comprimento de onda correspondente à cor do objeto é encontrado no lado oposto ao comprimento de onda da absorção máxima.

Figura 1

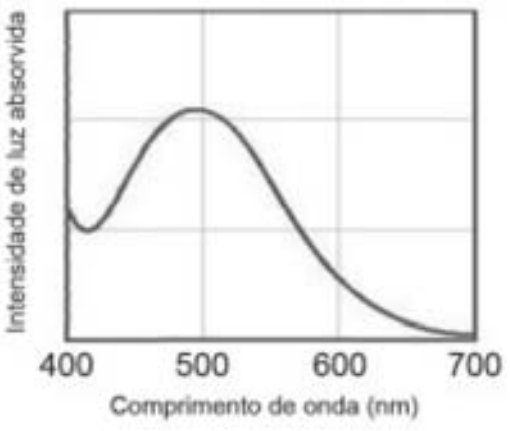

Figura 2

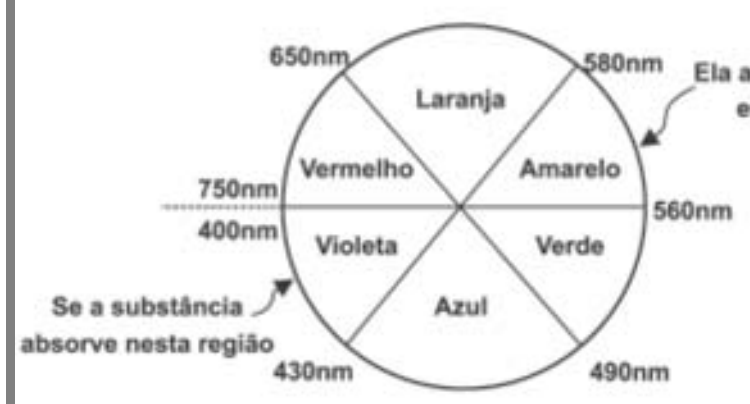

Brown, T. **Química a Ciência Central**. 2005 (adaptado).

Qual a cor da substância que deu origem ao espectro da Figura 1?

A) Azul.

B) Verde.

C) Violeta.

D) Laranja.

E) Vermelho.

Uma outra classificação foi a de itens como quantitativos, semiquantitativos e qualitativos. Quantitativos são os itens que necessitam obrigatoriamente de cálculo para a resolução, semiquantitativos são os que podem ser resolvidos pela análise de proporcionalidade (relações como "maior que", "menor que", "igual"), e qualitativos são os itens cuja solução prescinde da utilização de raciocínio ou relação matemática, sendo apenas conceituais. No Quadro 4, apresenta-se o exemplo de uma questão classificada como semiquantitativa (Questão 77 da Prova Azul de 2011). No Quadro 5, uma questão classificada como qualitativa (Questão 78 da Prova Azul de 2010).

**Quadro 4** - Questão 77, Prova Azul, 2011 – uma questão semiquantitativa

**QUESTÃO 77**

Para medir o tempo de reação de uma pessoa, pode-se realizar a seguinte experiência:

**I.** Mantenha uma régua (com cerca de 30 cm) suspensa verticalmente, segurando-a pela extremidade superior, de modo que o zero da régua esteja situado na extremidade inferior.

**II.** A pessoa deve colocar os dedos de sua mão, em forma de pinça, próximos do zero da régua, sem tocá-la.

**III.** Sem aviso prévio, a pessoa que estiver segurando a régua deve soltá-la. A outra pessoa deve procurar segurá-la o mais rapidamente possível e observar a posição onde conseguiu segurar a régua, isto é, a distância que ela percorre durante a queda.

O quadro seguinte mostra a posição em que três pessoas conseguiram segurar a régua e os respectivos tempos de reação.

| Distância percorrida pela régua durante a queda (metro) | Tempo de reação (segundo) |
|---|---|
| 0,30 | 0,24 |
| 0,15 | 0,17 |
| 0,10 | 0,14 |

Disponível em: http://br.geocities.com. Acesso em: 1 fev. 2009.

A distância percorrida pela régua aumenta mais rapidamente que o tempo de reação porque a

(A) energia mecânica da régua aumenta, o que a faz cair mais rápido.
(B) resistência do ar aumenta, o que faz a régua cair com menor velocidade.
(C) aceleração de queda da régua varia, o que provoca um movimento acelerado.
(D) força peso da régua tem valor constante, o que gera um movimento acelerado.
(E) velocidade da régua é constante, o que provoca uma passagem linear de tempo.

**Quadro 5.** Questão 78, Prova Azul, 2010/2 – uma questão qualitativa

**Questão 78**

Duas irmãs que dividem o mesmo quarto de estudos combinaram de comprar duas caixas com tampas para guardarem seus pertences dentro de suas caixas, evitando, assim, a bagunça sobre a mesa de estudos. Uma delas comprou uma metálica, e a outra, uma caixa de madeira de área e espessura lateral diferentes, para facilitar a identificação. Um dia as meninas foram estudar para a prova de Física e, ao se acomodarem na mesa de estudos, guardaram seus celulares ligados dentro de suas caixas. Ao longo desse dia, uma delas recebeu ligações telefônicas, enquanto os amigos da outra tentavam ligar e recebiam a mensagem de que o celular estava fora da área de cobertura ou desligado.
Para explicar essa situação, um físico deveria afirmar que o material da caixa, cujo telefone celular não recebeu as ligações é de

A madeira, e o telefone não funcionava porque a madeira não é um bom condutor de eletricidade.
B metal, e o telefone não funcionava devido à blindagem eletrostática que o metal proporcionava.
C metal, e o telefone não funcionava porque o metal refletia todo tipo de radiação que nele incidia.
D metal, e o telefone não funcionava porque a área lateral da caixa de metal era maior.
E madeira, e o telefone não funcionava porque a espessura desta caixa era maior que a espessura da caixa de metal.

Para a análise quantitativa, preparou-se o banco de dados do ENEM 2009 (a prova aplicada) de acordo com os microdados disponíveis (em janeiro de 2012) na página do INEP.

Esses microdados foram estruturados e validados. Tomou-se como referência a prova azul (foi feita a conversão das provas das demais cores para corresponder à numeração da prova azul), e foram separados dois grupos: todos os participantes (cerca de 4,3 milhões de estudantes) e os autodeclarados concluintes em 2009 (cerca de 940 mil).

Com o banco de dados preparado, foi feita a análise descritiva simples dos resultados, com a comparação entre as classificações feitas e os percentuais de acertos nas questões.

O levantamento do percentual de acerto de cada uma das questões de Física da prova de Ciências da Natureza foi elaborado, e esses resultados foram agrupados de forma simples entre as várias classificações feitas para os tipos de questão. No Anexo 1, apresenta-se o quadro das respostas dos estudantes concluintes em 2009 e de todos os estudantes nas questões de Física da prova de Ciências da Natureza do ENEM 2009 [3].

## Resultados: as Características das Provas segundo Elementos da Matriz de Referência

Há uma regularidade nas provas analisadas: cerca de um terço da prova de Ciências da Natureza, com 45 itens, é composta de questões de Física. Na Figura 1, apresentamos um histograma do número de questões em cada uma das provas (2009/1, a prova de 2009 que não foi aplicada pelo vazamento; 2009/2, a prova aplicada; 2010/1, a primeira prova aplicada; 2010/2, a prova aplicada extraordinariamente por defeitos na primeira; 2011).

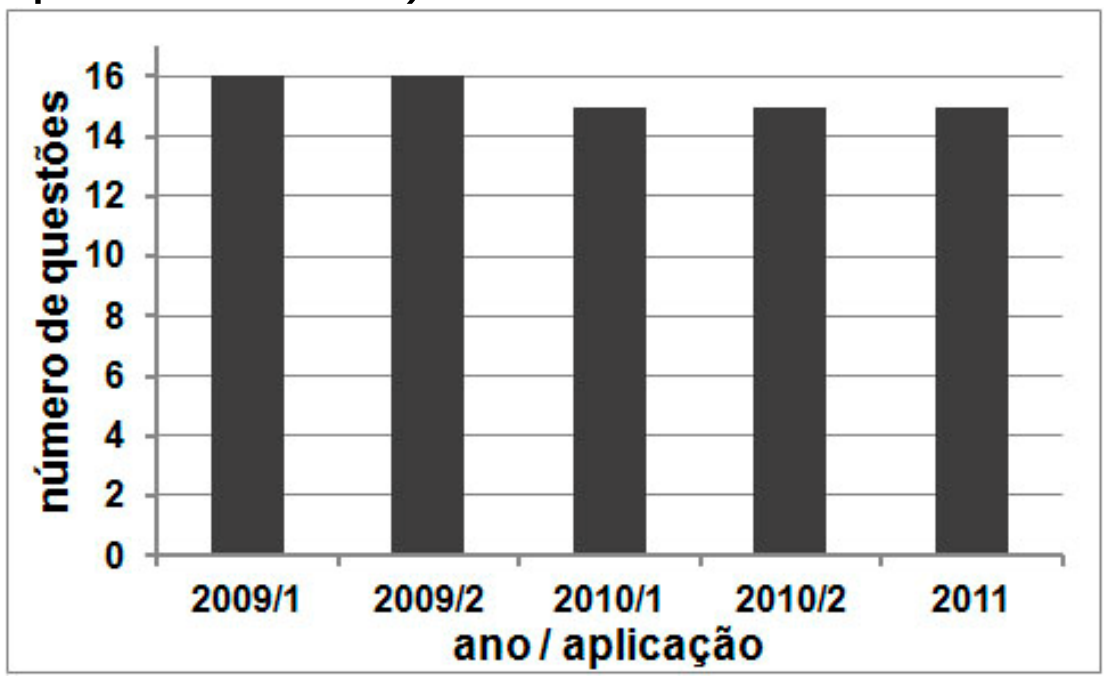

**Figura 1.** O número de questões de física nas provas de Ciências da Natureza do ENEM.

Na área de Ciências da Natureza, são propostas 8 competências, apresentadas no Quadro 1, distribuídas em 30 habilidades [9]. Há, como pode ser observado da Figura 2, uma concentração de questões na competência 6, "*Apropriar-se de conhecimentos da física para, em situações problema, interpretar, avaliar ou planejar intervenções científico-tecnológicas*", que é específica de conteúdos disciplinares de Física.

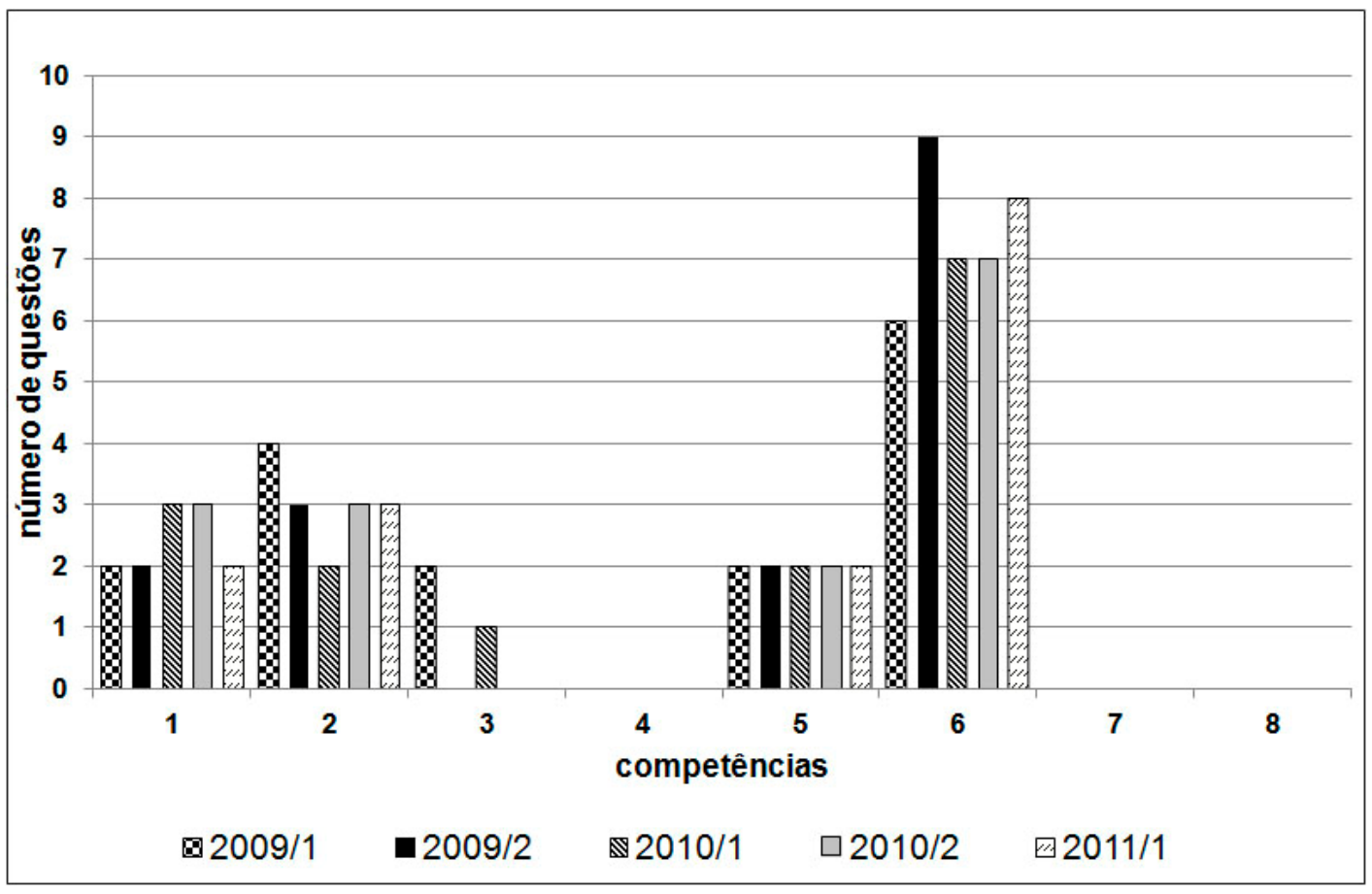

**Figura 2**. Distribuição das competências nas provas do ENEM.

A distribuição das habilidades pelos itens das provas indica que nem todas as 16 habilidades de Física são solicitadas em todas as provas; de fato, a construção da prova não é baseada estritamente em considerações disciplinares. Na Tabela 1, apresenta-se um quadro das habilidades contempladas nos itens da competência de área 6.

| Competência | Habilidade | 2009/1 | 2009/2 | 2010/1 | 2010/2 | 2011/1 |
|---|---|---|---|---|---|---|
| 6 | 20 | 1 | 2 | 1 | 2 | 3 |
| | 21 | 1 | 4 | 2 | 2 | 1 |
| | 22 | 2 | 2 | 1 | 1 | 1 |
| | 23 | 2 | 1 | 3 | 2 | 3 |

**Tabela 1.** Distribuição das habilidades na competência de área 6.

Também percebe-se que vários itens se encaixam em mais de uma habilidade. Isso se torna um problema quando se leva em conta que a Teoria de Resposta ao Item [10] pressupõe que o item deve ser unidimensional, isto é, contemplar somente um elemento ou célula da Matriz de Referência.

Os denominados "objetos de conhecimento" (ou conteúdos) na Matriz de Referência de Ciências da Natureza e suas Tecnologias são divididos em 7 grandes áreas, como apresentado no Quadro 2.

Nas provas de 2009 a 2011, a distribuição dos objetos de conhecimento pelas questões de física está apresentada na forma de gráficos na Figura 3.

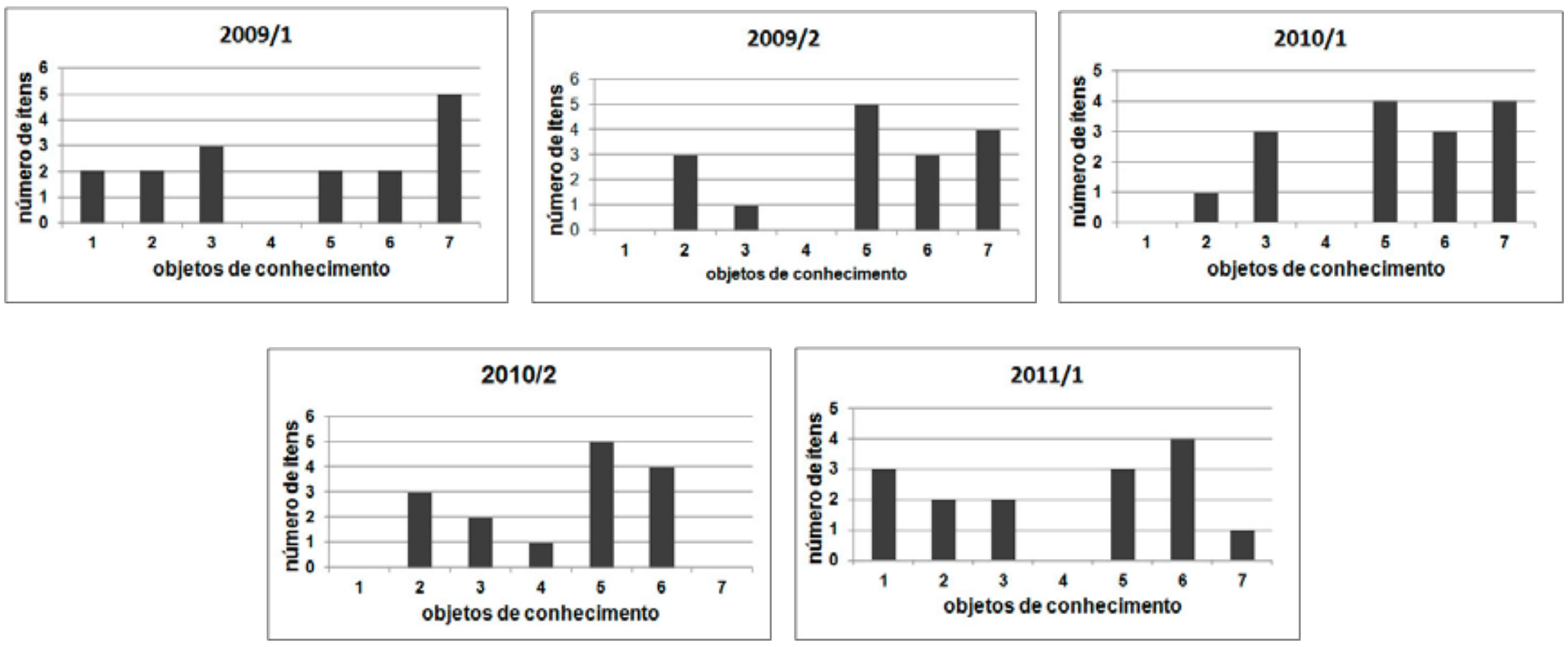

**Figura 3.** Número de itens por objeto de conhecimento em cada prova.

Observa-se alguma regularidade na distribuição dos conteúdos, sendo privilegiadas as áreas 5 (eletromagnetismo), 6 (oscilações e ondas) e 7 (física térmica), com mais de 50% das questões de cada prova. Os conteúdos de Mecânica (áreas 2, 3 e 4) correspondem a menos de 30% das questões; os conhecimentos da área 2 (dinâmica, entre outros) são avaliados em menos de 20% dos itens. Os itens classificados na área 3 em sua maioria compreendem questões genéricas de transformação de energia, no conteúdo específico “conceituação de trabalho, energia e potência”. Verifica-se que a maior parte do conteúdo de mecânica tem sido pouco abordado nas provas.

A discussão da distribuição dos conteúdos nos itens de física torna-se fundamental a partir do momento em que o ENEM passa a ser a única forma de acesso a muitas das universidades públicas brasileiras e, portanto, torna-se referência para a determinação do currículo ensinado nas escolas. Das provas até aqui aplicadas, observa-se que a Mecânica passa a ser um conhecimento menos exigido, e provavelmente menos ensinada. A Matriz de Referência portanto pode, sem muitas discussões, promover mudanças profundas nos currículos do Ensino Médio no Brasil.

## Resultados: as Características das Provas

O grande número (45 por área) de itens nas provas está relacionado às necessidades geradas pela utilização da Teoria da Resposta ao Item (TRI). Os candidatos dispõem, em média, de 3 minutos para a solução de cada item. Os textos-base dos itens possuem uma média de 10 (± 4) linhas de extensão; além do texto base, o enunciado também deve ser lido (em 2009, o número médio foi de 5 linhas), bem como as alternativas de resposta. Este tamanho é grande para o tempo médio disponível.

Na Figura 4, apresenta-se a classificação do nível de contextualização dos itens [12, 13]. Nesta classificação, um item tem alta contextualização se no texto estão presentes as informações relevantes para a sua solução, isto é, se a extração e processamento da informação contida no texto é necessária para a resolução do item; por outro lado, um baixo nível de contextualização está presente se a informação fornecida no texto não é essencial para responder ao item. A identificação do nível de contextualização do texto-base foi então classificada como pré-texto, onde ele poderia não estar presente para a adequada resolução do item [14], baixa contextualização, quando as informações necessárias para solução do item encontram-se em de 10% a 40% do número total de linhas do texto, média contextualização, quando a porcentagem sobe para entre 50% a 70%, e alta contextualização, quando o mais de 80% das linhas são necessárias para a resolução do item.

Da Figura 4, verifica-se que as provas mais recentes têm nível de contextualização (tal como definido por Nentwig [13]) mais altos.

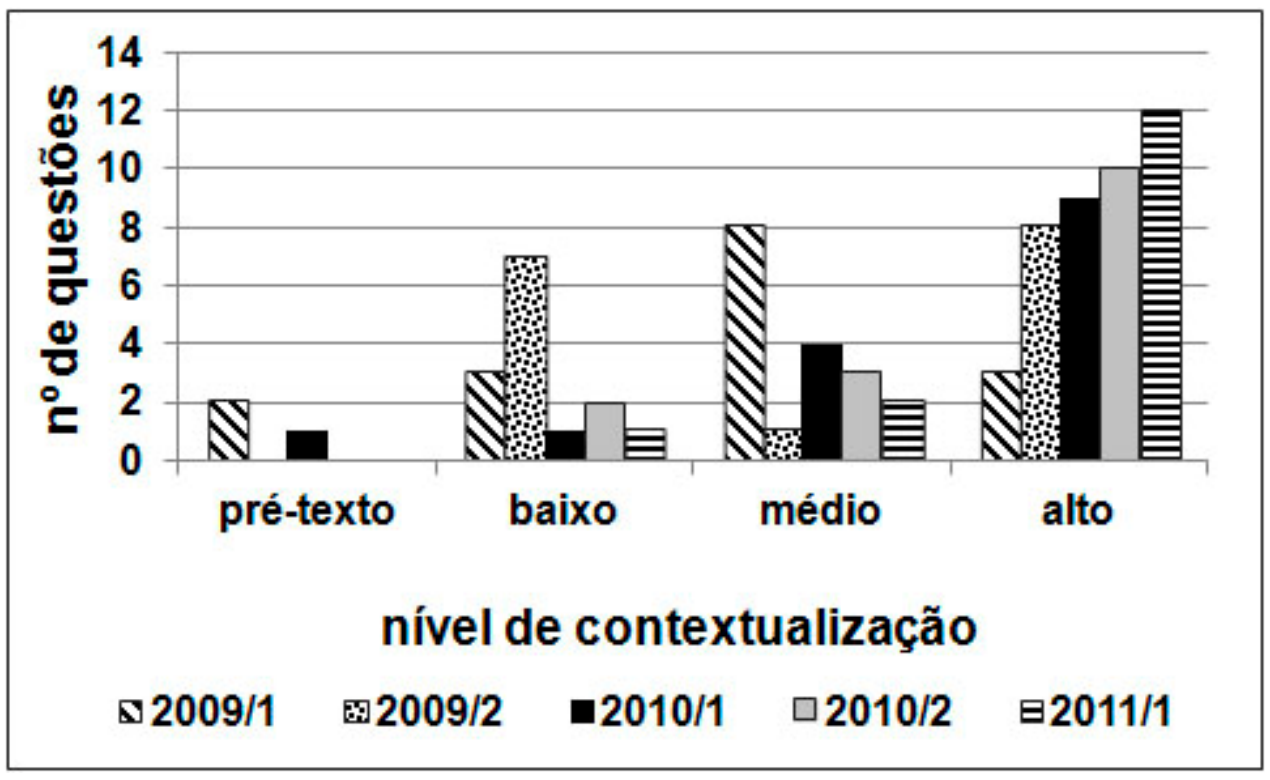

**Figura 4.** Nível de contextualização do itens por ano/aplicação.

Há itens em que não há exigência de algum conhecimento específico de física para resolução das questões, como mostrado na Figura 5. Cabe a reflexão sobre se após a reformulação do ENEM é importante que ele avalie, em algum nível, o conhecimento disciplinar dos alunos.

É importante ressaltar que não se pretende com este trabalho fazer qualquer discussão relativa à precisão conceitual das questões destas provas.

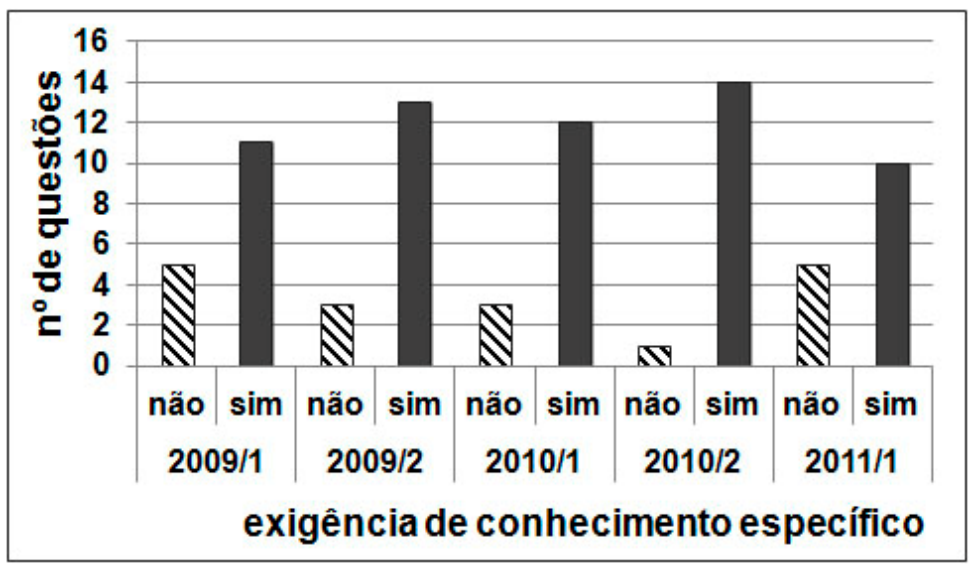

**Figura 5.** Exigência de cobrança de conhecimento específico de física.

Na Figura 6, são apresentados os percentuais das questões de Física, em cada prova, que são classificadas como quantitativas, qualitativas ou semiquantitativas. Desta Figura 6, observa-se que os itens da prova de física são predominantemente qualitativos, não exigindo raciocínios quantitativos ou avaliação da capacidade de resolução de problemas.

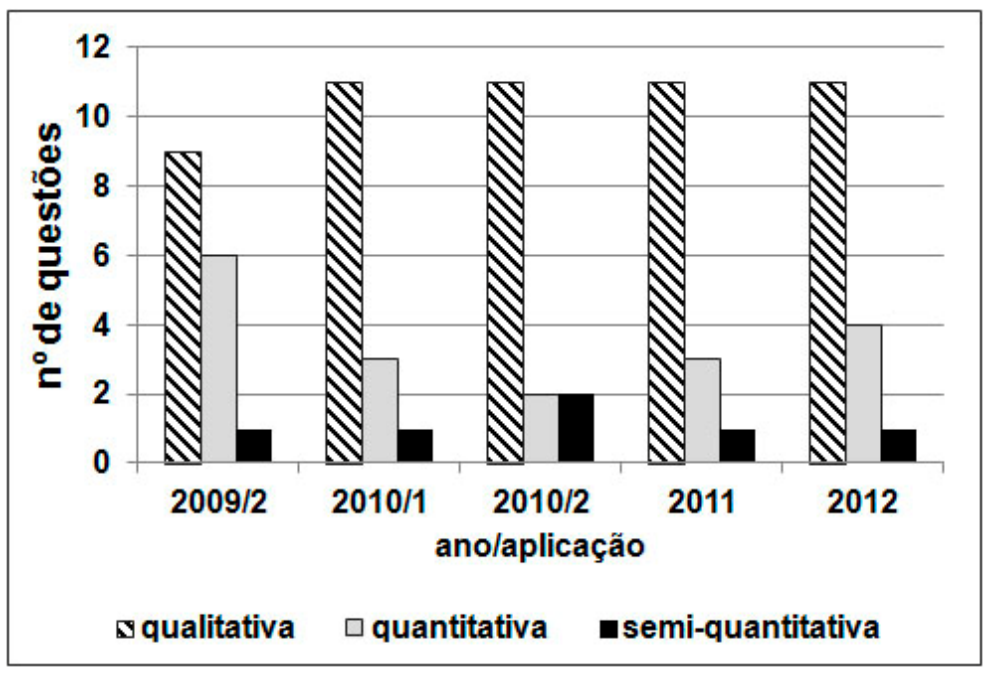

**Figura 6**. A distribuição por caráter quantitativo ou qualitativo das questões.

**Resultados: o Desempenho dos Alunos**

Os microdados do ENEM 2009 foram disponibilizados pelo INEP, permitindo a montagem de um banco de dados com as respostas dos candidatos por questão. A prova de referência para as análises é a prova azul, de Ciências da Natureza. Foram eliminados os faltosos, e separaram-se todos os participantes (2.555.594) e os auto-declarados concluintes em 2009 (944.162), estudantes do terceiro ano do ensino médio em 2009.

Na Figura 7, apresenta-se o percentual de acertos em cada uma das questões da prova aplicada de Ciências da Natureza de 2009, anteriormente citada como 2009/2. Em destaque, estão as questões de Física desta prova.

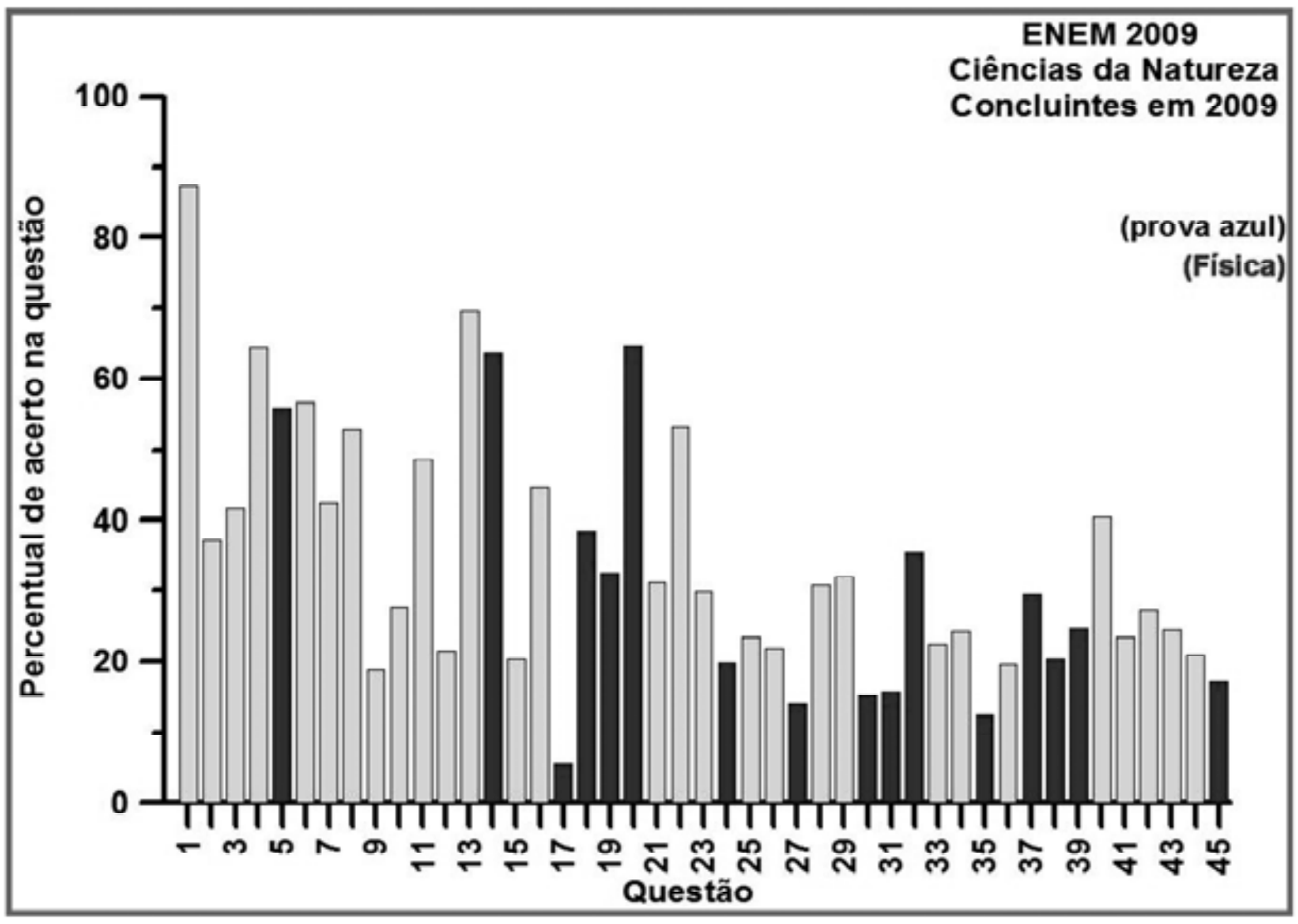

**Figura 7.** Percentual de acertos nas questões de Ciências da Natureza do ENEM 2009 – concluintes em 2009

Na Tabela 2, está o percentual de acertos entre os concluintes e entre todos os participantes do ENEM 2009, divididos entre as questões de Física classificadas como qualitativas e quantitativas. O desempenho, em média, é inferior nas questões que exigem algum tipo de raciocínio matemático.

| Itens qualitativos | | |
|---|---|---|
| Item | Concluintes | Todos |
| 5 | 55.9 | 57.5 |
| 14 | 63.6 | 68 |
| 18 | 38.5 | 39.5 |
| 20 | 64.6 | 65.5 |
| 24 | 19.7 | 19.4 |
| 27 | 14.0 | 14.1 |

| Itens qualitativos | | |
|---|---|---|
| Item | Concluintes | Todos |
| 31 | 15.8 | 15.4 |
| 32 | 35.5 | 35.5 |
| 37 | 29.5 | 30.7 |
| 39 | 24.7 | 22.5 |
| 45 | 17.2 | 17.1 |
| MÉDIA | 34.5 | 35.0 |

| Itens quantitativos | | |
|---|---|---|
| Item | Concluintes | Todos |
| **17** | 5.7 | 6.4 |
| **19** | 32.6 | 34.4 |
| **30** | 15.2 | 15.4 |
| **35** | 12.4 | 12.8 |
| **38** | 20.3 | 20.8 |
| MÉDIA | 17.2 | 18.0 |

**Tabela 2.** O percentual de acerto nas questões de física do ENEM 2009.

Na Tabela 3, está o percentual de acerto nas questões classificadas segundo a exigência de conhecimento específico. Novamente, o desempenho revela-se em média mais fraco nas questões que exigem conhecimento de física.

| Sem exigência de conhecimento específico | | |
|---|---|---|
| Item | Concluintes | Todos |
| 5 | 55.9 | 57.5 |
| 19 | 32.6 | 34.4 |
| 20 | 64.6 | 65.5 |
| **MÉDIA** | **51** | **52.5** |

| Exigência de conhecimento específico | | | Exigência de conhecimento específico | | |
|---|---|---|---|---|---|
| Item | Concluintes | Todos | Item | Concluintes | Todos |
| 14 | 63.6 | 68 | 31 | 15.8 | 15.4 |
| 17 | 5.7 | 6.4 | 32 | 35.5 | 35.5 |
| 18 | 38.5 | 39.5 | 35 | 12.4 | 12.8 |
| 19 | 32.6 | 34.4 | 37 | 29.5 | 30.7 |
| 24 | 19.7 | 19.4 | 38 | 20.3 | 20.8 |
| 27 | 14.0 | 14.1 | 39 | 24.7 | 22.5 |
| 30 | 15.2 | 15.4 | 45 | 17.2 | 17.1 |
| | | | **MÉDIA** | **24.6** | **25.1** |

**Tabela 3.** O percentual de acerto nas questões de física do ENEM 2009.

As respostas por alternativa nas questões de Física da prova de Ciências da Natureza do ENEM 2009 estão apresentadas em detalhes na parte 1 do Anexo, e na parte 2 deste Anexo estão os percentuais de acerto em todas as questões da prova de Ciências da Natureza. Esses resultados são apresentados para todos os participantes da prova de 2009 (4,18 milhões de candidatos inscritos, e 2,56 milhões de candidatos participando da prova) e para os auto-declarados concluintes do ensino médio em 2009 (940 mil candidatos).

## Discussão Final

Neste trabalho, apresentou-se uma análise das características qualitativas das questões do ENEM, com base em categorias associadas à adequação do exame aos objetivos propostos, ao tempo de reflexão na resolução da prova, o tipo de contextualização proposta no item e a utilização de procedimentos de raciocínio menos elementares e mais abstratos.

Fez-se também uma comparação entre os resultados do desempenho dos estudantes participantes do ENEM 2009 e o dos concluintes do ensino médio neste ano. Esses resultados foram obtidos a partir dos dados disponibilizados pelo INEP.

O perfil desse exame que emerge da análise é que ele se caracteriza por apresentar itens predominantemente qualitativos, extensos para o tempo disponível para resolvê-los e, de uma forma geral, bem contextualizados.

Além da extensão dos itens, dois outros pontos se destacam: o fato de que os itens podem representar mais de uma habilidade, ferindo assim o postulado da unidimensionalidade da Teoria de Resposta ao Item; e a distribuição não homogênea dos itens nos objetos específicos de conteúdo, privilegiando os temas que abordam conceitos de eletricidade e termodinâmica. Este fator, a médio e longo prazo, pode produzir sérias distorções no currículo de física do ensino médio no Brasil.

A avaliação aponta para a necessidade de reformulação da Matriz de Referência do ENEM, para que os objetos de conhecimento estejam mais alinhados com os apresentados no ensino médio, para que a distribuição de habilidades e competências se dê de forma mais homogênea, permitindo que o exame se torne um instrumento de maior utilidade para compor indicadores da qualidade da educação no ensino médio do país.

A partir dos resultados dos estudantes, observa-se preliminarmente que o percentual de acertos entre concluintes do ensino médio e candidatos em geral foi praticamente o mesmo para todas as questões da prova de Ciências da Natureza. Percebe-se diferença nas médias dos desempenhos em questões que envolvem análises quantitativas (com percentuais de acerto menores), na média dos desempenhos quando há exigência de conhecimento específico de física (mais baixos). Também, de forma um pouco inesperada, os itens de Física Térmica tiveram em média um percentual de acerto bastante pequeno.

Avaliações em larga escala visam fornecer dados para implementação, manutenção e reformulação de políticas públicas e que avaliações governamentais de larga escala colocam muitas vezes em cheque o professor em relação às suas práticas avaliativas. Portanto, compreender, conhecer e refletir sobre as características destas avaliações torna-se primordial para o professor.

## Agradecimentos

Este trabalho foi desenvolvido com apoio da CAPES/MEC, Programa Observatório da Educação 2010, Projeto “Avaliações Educacionais e o Ensino de Ciências e Matemática”.
Agradecemos a Marcelo Shoey de Oliveira Massunaga e Gustavo Rubini, pelo auxílio na montagem do banco de dados e pelas discussões enriquecedoras sobre os resultados obtidos.

## Referências Bibliográficas

**Anexo 1 – O desempenho dos estudantes nas questões de Física da prova de Ciências da Natureza do ENEM 2009.**

**Parte 1 –** O percentual de respostas em cada uma das alternativas, para os estudantes concluintes e para todos os estudantes. O caractere entre parênteses corresponde ao gabarito do item. O número da questão corresponde à prova azul.

**Q05 (E)**

| | Concluintes (%) | Todos (%) |
|---|---|---|
| A | 17.2 | 16.1 |
| B | 16.2 | 15.8 |
| C | 5.9 | 6.1 |
| D | 4.4 | 4.2 |
| E | 55.9 | 57.5 |
| Branco | 0.1 | 0.2 |
| Inválido | 0.2 | 0.2 |
| Total | 100.0 | 100.0 |

**Q14 (E)**

| | Concluintes (%) | Todos (%) |
|---|---|---|
| A | 6.2 | 5.3 |
| B | 15.6 | 12.9 |
| C | 10.4 | 9.9 |
| D | 3.9 | 3.6 |
| E | 63.6 | 68.0 |
| Branco | 0.1 | 0.1 |
| Inválido | 0.2 | 0.2 |
| Total | 100.0 | 100.0 |

**Q17 (E)**

| | Concluintes (%) | Todos (%) |
|---|---|---|
| A | 22.8 | 22.4 |
| B | 27.4 | 26.3 |
| C | 25.1 | 25.1 |
| D | 18.6 | 19.5 |
| E | 5.7 | 6.4 |
| Branco | 0.3 | 0.3 |
| Inválido | 0.1 | 0.1 |
| Total | 100.0 | 100.0 |

**Q18 (E)**

| | Concluintes (%) | Todos (%) |
|---|---|---|
| A | 18.2 | 18.3 |
| B | 16.1 | 15.2 |
| C | 9.5 | 9.4 |
| D | 17.5 | 17.4 |
| E | 38.5 | 39.5 |
| Branco | 0.1 | 0.2 |
| Inválido | 0.1 | 0.1 |
| Total | 100.0 | 100.0 |

**Q19 (D)**

| | Concluintes (%) | Todos (%) |
|---|---|---|
| A | 21.6 | 20.8 |
| B | 14.2 | 13.5 |
| C | 19.9 | 19.8 |
| D | 32.6 | 34.4 |
| E | 11.3 | 11.1 |
| Branco | 0.2 | 0.2 |
| Inválido | 0.1 | 0.1 |
| Total | 100.0 | 100.0 |

**Q20 (E)**

| | Concluintes (%) | Todos (%) |
|---|---|---|
| A | 9.5 | 8.4 |
| B | 6.8 | 6.1 |
| C | 6.6 | 6.0 |
| D | 12.3 | 13.8 |
| E | 64.6 | 65.5 |
| Branco | 0.1 | 0.1 |
| Inválido | 0.1 | 0.1 |
| Total | 100.0 | 100.0 |

**Q24 (A)**

| | Concluintes (%) | Todos (%) |
|---|---|---|
| A | 19.7 | 19.4 |
| B | 18.1 | 17.4 |
| C | 22.3 | 23.0 |
| D | 31.9 | 32.7 |
| E | 7.9 | 7.3 |
| Branco | 0.1 | 0.1 |
| Inválido | 0.1 | 0.1 |
| Total | 100.0 | 100.0 |

**Q27 (D)**

| | Concluintes (%) | Todos (%) |
|---|---|---|
| A | 41.7 | 40.8 |
| B | 20.6 | 20.1 |
| C | 6.7 | 6.8 |
| D | 14.0 | 14.1 |
| E | 16.8 | 17.8 |
| Branco | 0.2 | 0.2 |
| Inválido | 0.2 | 0.2 |
| Total | 100.0 | 100.0 |

**Q30 (D)**

| | Concluintes (%) | Todos (%) |
|---|---|---|
| A | 21.8 | 22.8 |
| B | 27.6 | 25.7 |
| C | 24.3 | 23.9 |
| D | 15.2 | 15.4 |
| E | 10.6 | 11.9 |
| Branco | 0.3 | 0.3 |
| Inválido | 0.0 | 0.0 |
| Total | 100.0 | 100.0 |

**Q31 (E)**

| | Concluintes (%) | Todos (%) |
|---|---|---|
| A | 21.4 | 21.2 |
| B | 6.2 | 6.4 |
| C | 38.8 | 39.7 |
| D | 17.5 | 17.0 |
| E | 15.8 | 15.4 |
| Branco | 0.2 | 0.2 |
| Inválido | 0.1 | 0.1 |
| Total | 100.0 | 100.0 |

**Q32 (B)**

| | Concluintes (%) | Todos (%) |
|---|---|---|
| A | 19.2 | 18.8 |
| B | 35.5 | 35.5 |
| C | 24.0 | 24.1 |
| D | 15.2 | 15.7 |
| E | 5.7 | 5.5 |
| Branco | 0.3 | 0.3 |
| Inválido | 0.1 | 0.1 |
| Total | 100.0 | 100.0 |

**Q35 (A)**

| | Concluintes (%) | Todos (%) |
|---|---|---|
| A | 12.4 | 12.8 |
| B | 26.1 | 25.7 |
| C | 30.4 | 30.5 |
| D | 22.8 | 22.6 |
| E | 7.9 | 8.0 |
| Branco | 0.4 | 0.4 |
| Inválido | 0.1 | 0.1 |
| Total | 100.0 | 100.0 |

**Q37 (D)**

| | Concluintes (%) | Todos (%) |
|---|---|---|
| A | 20.5 | 21.6 |
| B | 17.4 | 15.8 |
| C | 9.1 | 8.6 |
| D | 29.5 | 30.7 |
| E | 23.1 | 22.9 |
| Branco | 0.2 | 0.2 |
| Inválido | 0.2 | 0.1 |
| Total | 100.0 | 100.0 |

**Q38 (D)**

| | Concluintes (%) | Todos (%) |
|---|---|---|
| A | 12.1 | 11.2 |
| B | 26.7 | 26.0 |
| C | 27.2 | 27.3 |
| D | 20.3 | 20.8 |
| E | 13.3 | 14.2 |
| Branco | 0.3 | 0.4 |
| Inválido | 0.1 | 0.1 |
| Total | 100.0 | 100.0 |

**Q39 (B)**

| | Concluintes (%) | Todos (%) |
|---|---|---|
| A | 45.3 | 49.2 |
| B | 24.7 | 22.5 |
| C | 13.1 | 12.5 |
| D | 10.0 | 9.7 |
| E | 6.6 | 5.7 |
| Branco | 0.2 | 0.2 |
| Inválido | 0.1 | 0.1 |
| Total | 100.0 | 100.0 |

**Q45 (B)**

| | Concluintes (%) | Todos (%) |
|---|---|---|
| A | 23.7 | 23.9 |
| B | 17.2 | 17.1 |
| C | 16.2 | 17.3 |
| D | 12.1 | 12.5 |
| E | 30.4 | 28.9 |
| Branco | 0.3 | 0.3 |
| Inválido | 0.1 | 0.1 |
| Total | 100.0 | 100.0 |

**Parte 2** – O percentual de acertos por questão (Prova Azul, 2009) da Prova de Ciências da Natureza para concluintes em 2009 (autodeclarados) e para todos os participantes.

| Questão | % acerto Concluintes | % acerto Todos |
|---|---|---|
| **1** | 87.3 | 89.1 |
| **2** | 37.1 | 39.3 |
| **3** | 41.7 | 43.2 |
| **4** | 64.4 | 64.5 |
| **5** | 55.9 | 57.5 |
| **6** | 56.7 | 57.5 |
| **7** | 42.3 | 46.1 |
| **8** | 52.9 | 54.7 |
| **9** | 18.9 | 18.1 |
| **10** | 27.7 | 28.7 |
| **11** | 48.5 | 49.0 |
| **12** | 21.6 | 20.8 |
| **13** | 69.6 | 74.8 |
| **14** | 63.6 | 68.0 |
| **15** | 20.3 | 19.3 |
| **16** | 44.8 | 46.8 |
| **17** | 5.7 | 6.4 |
| **18** | 38.5 | 39.5 |
| **19** | 32.6 | 34.4 |
| **20** | 64.6 | 65.5 |
| **21** | 31.2 | 30.8 |
| **22** | 53.3 | 59.0 |
| **23** | 29.9 | 34.5 |
| **24** | 19.7 | 19.4 |
| **25** | 23.4 | 22.3 |
| **26** | 22.0 | 21.8 |
| **27** | 14.0 | 14.1 |
| **28** | 30.7 | 31.3 |
| **29** | 31.9 | 31.5 |
| **30** | 15.2 | 15.4 |
| **31** | 15.8 | 15.4 |
| **32** | 35.5 | 35.5 |
| **33** | 22.5 | 20.7 |
| **34** | 24.2 | 25.4 |
| **35** | 12.4 | 12.8 |
| **36** | 19.6 | 19.4 |
| **37** | 29.5 | 30.7 |
| **38** | 20.3 | 20.8 |
| **39** | 24.7 | 22.5 |
| **40** | 40.4 | 41.1 |
| **41** | 23.4 | 23.6 |
| **42** | 27.4 | 28.5 |
| **43** | 24.5 | 24.8 |
| **44** | 20.9 | 20.8 |
| **45** | 17.2 | 17.1 |